# Optically-patterned nuclear doughnuts in GaAs/MnAs heterostructures

J. Stephens, J. Berezovsky, R. K. Kawakami, A. C. Gossard, and D. D. Awschalom

*Center for Spintronics and Quantum Computation, University of California, Santa Barbara, CA 93106*

## Abstract

We demonstrate a scheme for optically patterning nuclear spin polarization in semiconductor/ferromagnet heterostructures. A scanning time-resolved Kerr rotation microscope is used to image the nuclear spin polarization that results when GaAs/MnAs epilayers are illuminated with a focused laser having a Gaussian profile. Rather than tracking the intensity profile of the laser spot, these images reveal that the nuclear polarization forms an annular lateral structure having circular symmetry with a dip rather than a peak at its center.

The nascent field of spintronics has focused much attention on the potential for creating[1], manipulating[2], and detecting[3] spin polarized electrons in the solid state, particularly in semiconductor systems.[4] In addition, nuclear spin polarization has been shown[5] to be an important consideration and of potential utility for various spintronic devices and possibly quantum computing.[6] Recent work demonstrated a new technique dubbed Ferromagnetic Imprinting (FI) for optically generating highly non-equilibrium (up to ~20%) nuclear spin polarizations in ferromagnet-semiconductor heterostructures.[7] Because the level of nuclear polarization is strongly dependent on the illumination intensity, the FI process offers a route to producing spatially modulated nuclear spin polarizations simply by varying the intensity of the excitation in the lateral directions. Such optical patterning of nuclear spin polarization into arbitrary shapes could be achieved, for instance, by using Fourier optics to generate an arbitrary excitation beam profile. In addition, spins created by the FI process are polarized in the plane of the sample, a geometry that cannot be accessed simply by illumination with circularly polarized light.

In this Letter we present micron resolution spatial images of nuclear polarization in MnAs/n-GaAs epilayers produced by a stationary ~20 μm diameter laser beam using the FI process. Surprisingly, these images reveal a pronounced decrease in the polarization magnitude at the center of the Gaussian excitation spot where the intensity is highest. Further measurements indicate that the dip is caused in part by enhanced electron drift or diffusion away from the center of the excitation spot, yielding rings of polarized nuclei.

In the original study of the FI process ferromagnet/GaAs heterostructures were measured using overlapped ~40 μm diameter pump and probe beams to create and probe electron spins using the time resolved Faraday rotation technique.[7,8] In addition to observing electron spin dynamics, the photo-excited electrons created by the pump and probe pulses served to set up the nuclear spin polarization, an effect mediated by the ferromagnetic proximity polarization (FPP)[9] and subsequent dynamic nuclear polarization.[10] Here we report the unexpected micron-scale lateral structure in spatially inhomogeneous nuclear spin polarizations in similar samples using a three-beam geometry. Samples are composed of MBE-grown 15 nm type-A MnAs[11] films on top of 500 nm of n-doped (Si: 7e16 $cm^{-3}$) GaAs and a 400 nm $Al_{0.7}Ga_{0.3}As$ etch-stop layer with the underlying semi-insulating GaAs (100) substrate and buffer layers removed using a chemically selective wet etch.[12]

To generate a steady state nuclear spin polarization $\boldsymbol{I_n}$, we employ a linearly polarized "imprinting" beam from an 813 nm continuous wave diode laser which is focused onto the sample to a spot size of ~20 μm. The initially non-spin-polarized photo-excited electrons rapidly interact with the MnAs and become spin polarized antiparallel to the ferromagnet's magnetization *M*.[10] In an applied magnetic field $B_{app}$ = 0.21 T parallel to *M*, these electron spins couple to the nuclear spin system via the contact hyperfine interaction resulting in dynamic nuclear polarization (DNP) which reaches steady-state with a characteristic time constant of several minutes.

To locally probe the nuclear spin state, we measure the Larmor precession frequency of *electron* spins using a low-temperature time-resolved Kerr rotation (TRKR) microscope.[13] We create ~1 μm pump/probe spots which act only as small perturbations to the nuclear spin polarization prepared by the imprinting beam. The ~150 fs duration, 816 nm pump (450 μW) and probe (350 μW) pulses are generated by a mode-locked Ti:sapphire laser with a 76 MHz repetition rate. The circularly polarized pump beam excites electrons spin-polarized normal to the GaAs surface which precess at the Larmor frequency $\nu_L$ in the presence of a transverse magnetic field, $B_{app}$.[8] After a time Δt, the linearly polarized probe pulse is reflected from the GaAs and its polarization rotates due to the Kerr effect by an angle $\theta_K$, which is proportional to the component of electron spin normal to the sample. As we sweep Δt, $\theta_K$ oscillates at the Larmor frequency and decays with the transverse spin lifetime $T_2^*$ according to the relation $\theta_K = A \exp(-\Delta t/T_2^*) \cos(2\pi\nu_L\Delta t)$. The Larmor frequency is given by $\nu_L = g\mu_B(B_{app} + \boldsymbol{B_n})/h$ where g is the Landé g-factor, $\mu_B$ the Bohr magneton, h Plank's constant, and $B_n$ is the effective magnetic field due to the nuclear spin polarization (via the hyperfine coupling). By fitting $\theta_K(\Delta t)$ to obtain $\nu_L$ and assuming g = -0.44 for GaAs, we use the measured Larmor frequency as a local magnetometer of nuclear spin polarization since for GaAs, $\boldsymbol{I_n} \approx B_n(\mathrm{T})/5.3$.[14]

Figure 1a shows a plot of imprinting beam intensity versus position as measured by a charge coupled device camera as well as a linecut of nuclear polarization taken at T = 5 K and with the imprinting beam power set to 25 mW.[15] These plots show that it is indeed possible to setup local nuclear spin polarizations by spatially varying the intensity of the optical excitation. Two features in the nuclear polarization, however, stand out in contrast to the imprinting beam profile. First, rather than increasing monotonically towards the intense center of the imprinting beam, $B_n$ has two peaks at positions ~12 μm from the middle and then decreases toward the center to a value that is only ~2/3 of its maximum. Second, the full width at half maximum of nuclear polarization is ~5 times greater than that of the imprinting beam. This second characteristic is due at least in part to the motion of the FPP-oriented electron spins that are continually created by the imprinting beam. These spin-polarized electrons, which are responsible for the DNP, are able to spread (via diffusion or drift) for the duration of their longitudinal (parallel to $B_{app}$) spin lifetime $T_1$.[13] The two-dimensional image of nuclear polarization shown in Fig. 1b illustrates that $B_n$ has roughly the same circular symmetry as the imprinting beam.

Linecuts of $B_n$ versus lateral position for several imprinting beam powers ranging from 1 to 33 mW are plotted in Fig. 2a. The threshold power above which the nuclear spin polarization becomes appreciable occurs between 1 and 1.6 mW. In electrically biased structures, the onset of nuclear polarization was shown to be related to a lowering of the MnAs/GaAs Schottky barrier to flat-band conditions.[16] In our case, we believe that the photo-excited carriers screen the Schottky barrier, resulting in a similar turn-on of the nuclear polarization. Furthermore, we see that the dip in $B_n$ is present for all imprinting powers above 1.6 mW and that its relative depth increases monotonically with power. In contrast, we observe that while the maximum of the nuclear polarization initially increases with imprinting beam power, it decreases significantly at the highest intensities. The values of peak nuclear polarization and depth of the minima are shown for several powers in the rightmost panels of Fig. 2a. Finally, it is worth noting that the nuclear polarization and imprinting beam profiles are much different even for the relatively modest power of 1.6 mW.

Figure 2b shows the temperature dependence of $B_n$ with the imprinting beam set to 18 mW.[17] Here we see a decrease in both overall nuclear polarization[7] as well as the depth of the dip from T = 6-45 K, afterwhich the dip has disappeared (right hand panels of Fig. 2b). In addition, the full-width at half maximum of $B_n$ decreases from 68 μm to 43 μm over this temperature range. Similar behavior was recently reported for lithographically patterned structures and may be related to a decreasing diffusion distance due to the suppression of electron $T_1$ time with increasing temperature.[13]

Numerous effects may give rise to the observed nuclear polarization profile. Spin relaxation as well as drift and diffusion of the photo-excited electrons are important factors. In addition, local heating caused by the imprinting beam can create a temperature gradient. Such a gradient would enhance the mobility[13] (and hence the diffusivity) of electrons near the central, warmer region. Higher temperatures are also accompanied by a decrease in electron spin lifetime, and would result in a lower average electron and nuclear spin polarization in the central region. The transverse electron spin lifetime $T_2^*$ is also modified significantly by dephasing due to the inhomogeneous effective nuclear field. Figure 2c shows a linecut of the nuclear polarization overlaid with the effective transverse spin lifetime. The contribution from inhomogeneous dephasing follows the shape of the absolute value of the derivative of the nuclear polarization. However, this cannot explain the central minimum. Instead, it is most likely due to high carrier concentration or local heating in the central region.

To explore these possibilities, asymmetry was introduced to the nuclear polarization linecut measurements by spatially separating the previously overlapped pump and probe spots by 6 microns. In this case, probe pulses interact with pump spins which have moved (via drift or diffusion) ~6 μm towards the probe. The TRKR signal is therefore small at $\Delta t = 0$, initially increases for $\Delta t > 0$ as the pump spins arrive, and finally decreases for longer Δt as the spins decay due to spin relaxation and further motion away from the probe. This behavior is illustrated in Fig. 3a. Figures 3b, 3c, and 3d show intensity plots of TRKR signal versus delay Δt and position for three different cases, respectively: pump to the right of the probe; pump to the left of the probe; and pump and probe overlapped. In Fig. 3b, at the positions of peak nuclear polarization, a subtle difference in the amplitude of the TRKR signal is evident as highlighted by the dashed boxes. In this case, the amplitude at the right maximum is larger than at the left. When the orientation of the pump and probe spots is reversed as in Fig. 3c, the amplitude at the left maximum is greater. In the case where the pump and probe spots are overlapped the asymmetry is gone, as shown in Fig. 3d. These data suggest that electron spins are driven radially outward from the central region of the imprinting beam. In Fig. 3c for instance, where the pump spins are created to the right of the probe spot, the pump spins are pushed left towards the probe spot when measuring on the left side of the dip, resulting in an enhancement of the TRKR signal. Conversely, when measuring on the right side of the dip, the pump spins are pushed away from the probe spot, decreasing the TRKR signal. These spins move outward rapidly and thus have a decreased likelihood of undergoing spin flips with nuclei in the center.

We have demonstrated that nuclear spin polarization in ferromagnet/semiconductor heterostructures can be patterned by spatially modulating the excitation intensity. In the case of an intense Gaussian beam, the resulting nuclear polarization does not match the profile of the imprinting beam but rather forms an annular ring. We acknowledge support from DARPA, DMEA, and AFOSR.

**Figure Captions**

Fig. 1. (a) Nuclear polarization (solid line) created with an imprinting beam power of 25 mW shows a dramatic dip in the center and much wider profile compared to the imprinting beam (dashed line). (b) Two dimensional image of nuclear polarization shows the roughly circular symmetry which results from the Gaussian imprinting beam. All nuclear polarization linecuts are taken through the center of the doughnut as depicted by the dashed arrow.

Fig. 2. (a) Intensity dependence (T = 5 K, $B_{app}$ = 0.21 T) illustrates the non-monotonic behavior of the nuclear polarization with increasing imprinting beam power. The right panels show the peak polarization (top) and depth of the central minima in nuclear polarization (bottom) for several imprinting beam powers. (b) Temperature dependence of nuclear polarization ($B_{app}$ = 0.21 T, P = 18 mW). The peak polarization and depth of the central minima are shown in the top and bottom right panels, respectively. (c) Electron effective transverse spin lifetime $T_2^*$ and nuclear spin polarization vs. position (T = 5 K, $B_{app}$ = 0.21 T, P = 18 mW).

Fig. 3. (a) Time-resolved Kerr rotation signal at a fixed position with the pump and probe spots separated by 6 μm. The signal increases over the first ~1.5 ns as pump spins arrive under the probe spot and then decreases as they continue to move away and decay. (b-d) TRKR signal vs. position (relative to the imprinting beam) (T = 5 K, $B_{app}$ = 0.21 T) for the case of (b) pump to the left of the probe, (c) pump to the right of the probe, and

(d) pump and probe overlapped. The dashed boxes in (b) and (c) highlight the difference in TRKR signal intensity between left and right of the center dip in nuclear polarization.

[17] The temperature is measured by a sensor mounted on the copper cold finger several millimeters from the sample.  Local heating due to the imprinting beam is almost certainly present.

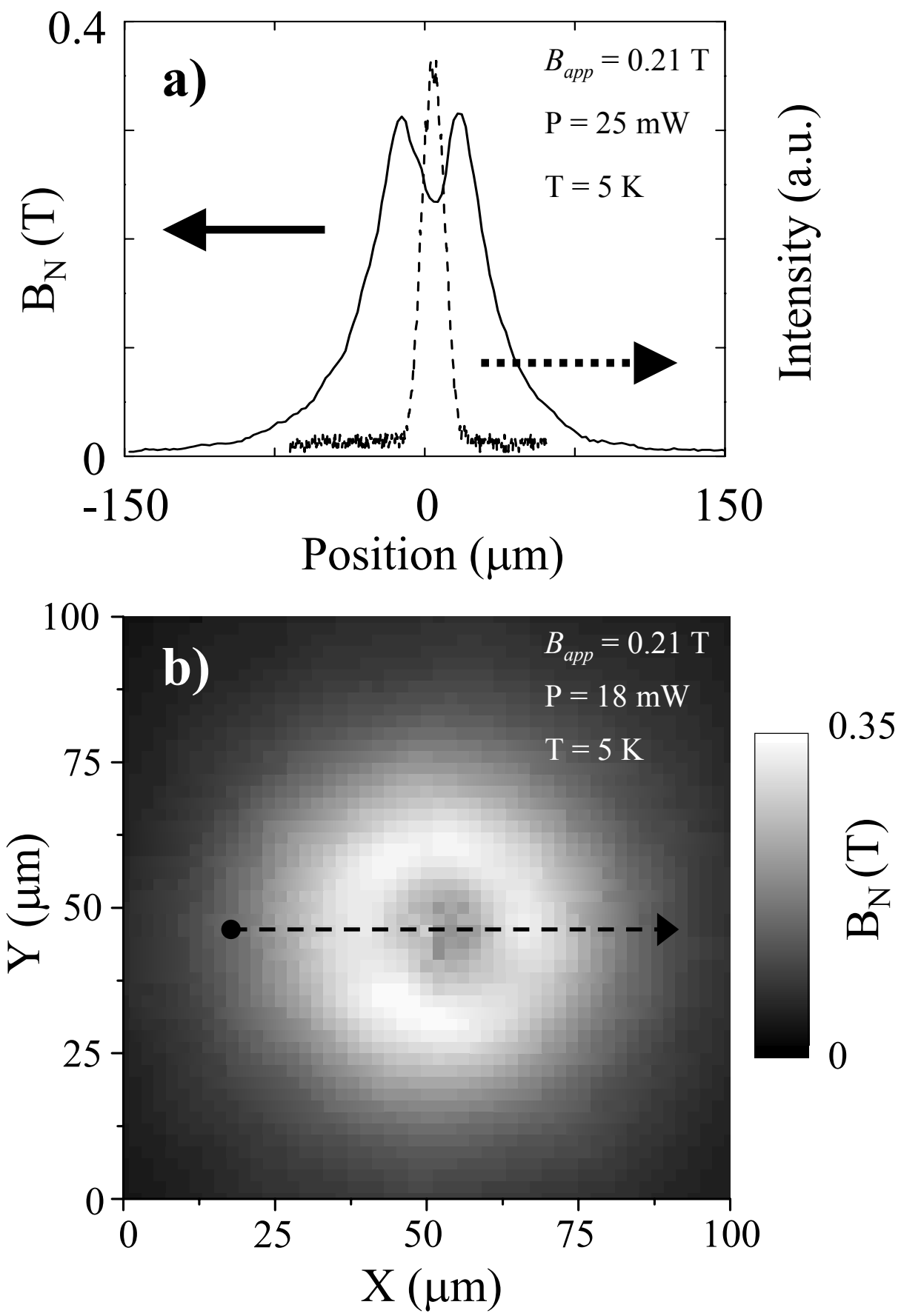

Figure 1

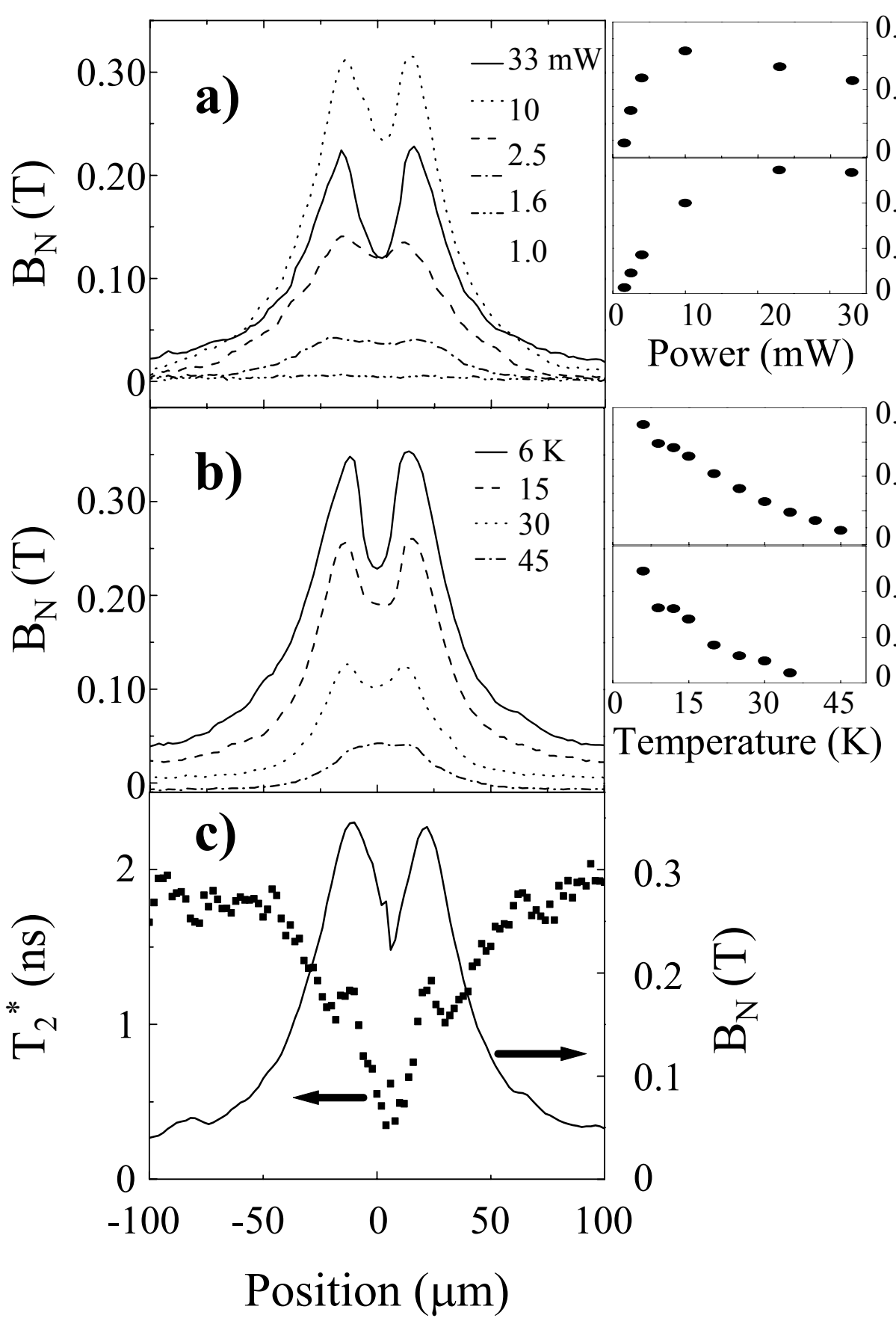

Figure 2

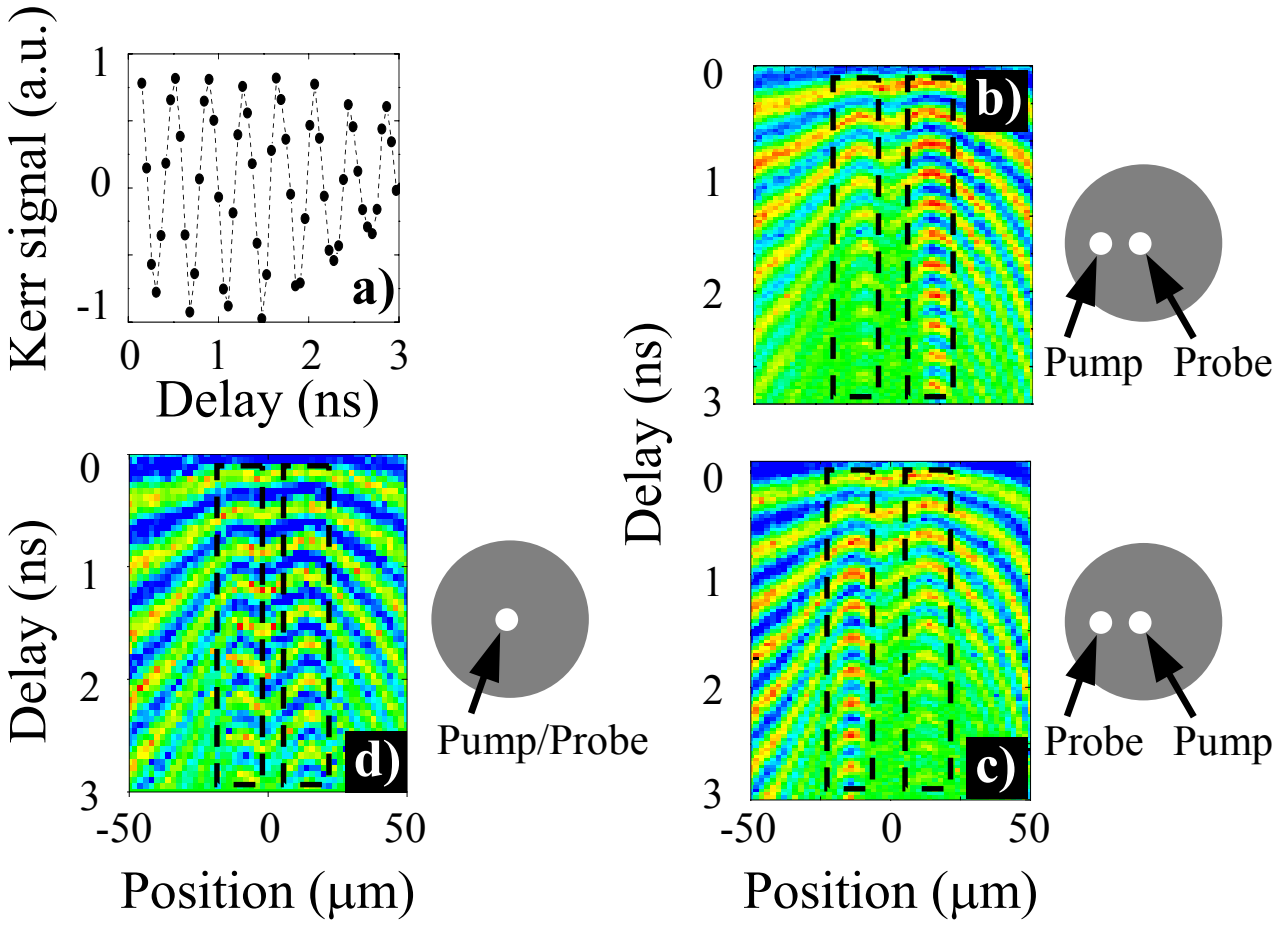

Figure 3